# 3D printing by two-photon polymerization of hollow microneedles for interstitial fluid extraction

Tiago Elias Abi-Ramia Silva,[1] Stephan Kohler,[1] Nicolas Bartzsch,[1]

Felix Beuschlein,[2,3,4] and Andreas T. Güntner[1,2,*]

[1]*Human-centered Sensing Laboratory, Department of Mechanical and Processing Engineering, ETH Zurich, CH-8092 Zurich, Switzerland*

[2]*Department of Endocrinology, Diabetology, and Clinical Nutrition, University Hospital Zurich (USZ) and University of Zürich (UZH), CH-8091 Zurich, Switzerland*

[3]*Medizinische Klinik und Poliklinik IV, Klinikum der Universität, Ludwig-Maximilians-Universität, Munich, Germany*

[4]*The LOOP Zurich – Medical Research Center, Zurich, Switzerland*

*Corresponding author.
E-mail address: andregue@ethz.ch

# Abstract

Dermal interstitial fluid (ISF) is a rich source of biomarkers (e.g., glucose) that can be used for continuous health monitoring with wearable sensors. Hollow microneedle devices are a promising solution to extract ISF on demand by penetrating the skin with minimal pain. However, they rely on inserting bio-incompatible materials (e.g., silicon) into individuals, limiting the application time. Here, the direct 3D printing of polymer hollow microneedles on silicon-based microfluidic devices and the successful in-vivo extraction of ISF are demonstrated. Our additive manufacturing approach enables the versatile combination of materials and rapid prototyping of microneedle geometry. After improving the design through finite element modeling, a hollow microneedle geometry was printed by two-photon polymerization and experimentally characterized with mechanical and fluidic tests. Microneedles were fabricated with high accuracy (i.e., 997 ± 2 µm) and reliably interfaced with the microfluidic chip (i.e., centerline alignment within 5% of diameter). The needles demonstrated sufficient mechanical strength (i.e., 411 ± 3 mN per needle) to endure at least 10 consecutive insertions into simulated skin. Biocompatibility and ISF extraction were demonstrated in an in-vivo 72-hour test, showing the safety and reliability of our approach. Such a platform is promising for minimally invasive, continuous monitoring of biomarkers in ISF, aiding in medical diagnoses and personalized health treatments.

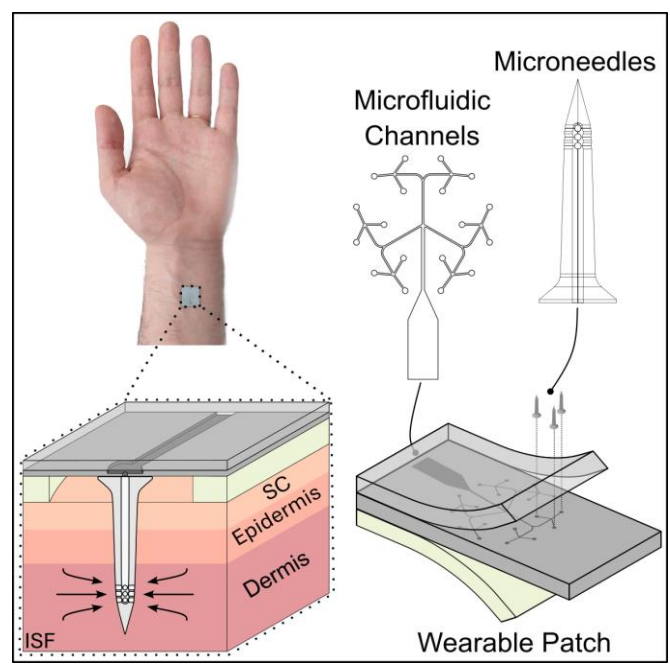

# 1. Introduction

The continuous availability of information on critical health parameters is a key bottleneck in realizing the envisioned P4 (predictive, preventive, personalized, and participatory) medicine [1]. However, biomarker sampling frequency is limited by medical practice traditionally relying on invasive and sometimes painful methods such as blood sampling with hypodermic needles [2]. While blood is a rich source of biomarkers, information can also be extracted from other biofluids, such as sweat [3], exhaled breath [4-6], and interstitial fluid (ISF). Dermal ISF has been suggested as a medium for medical analysis [7-9] as its biomarker composition is well correlated to that of blood for both the presence of chemical species [10, 11] and their temporal dynamics [12, 13].

Microneedles (MNs) have been widely proposed for ISF extraction [14] and biomarker sensing (e.g., glucose [13, 15] and electrolytes [16, 17]) due to their painless insertion, a feature that stems from their tips not reaching human skin's nerve endings [18]. This painless nature of MNs reassures patient comfort and acceptance [19]. Most established are non-hollow MNs that can be coated with sensing layers for on-site molecular analysis [8]. This simplicity, however, restricts sensing materials to those known to be biocompatible [20]. A promising alternative are hollow MNs. These can extract ISF on demand, allowing it to be analyzed outside the body, thus decoupling sampling from sensing. As a result, the sensing material and geometry can be chosen with fewer restrictions, and the sensing element can even be regenerated (e.g., by flushing) or calibrated between measurement cycles [21]. Hollow MNs also enable the development of highly integrated, compact, closed feedback-loop systems for drug delivery based on real-time health monitoring [22].

Most well-established fabrication methods of hollow MNs are silicon-based [23], generally limiting MN geometry to that produced through photolithography and etching [24, 25]. Further, despite its FDA approval for single-use MNs (e.g., NanoPass' MicronJet [26, 27]), silicon is not biocompatible for long-term use [28]. Also glass and metal have been used, but their geometry is usually restricted to conical, sharp outer shells with a hollow channel along their longitudinal axis [29]. More promising are polymer-based MNs due to their biocompatibility [30] and non-brittle behavior [31]. Ideally, such polymer-based MNs are directly interfaced with traditional microfluidic circuits to capitalize on the established knowledge for ISF manipulation and analysis [32], but this has not been demonstrated yet.

Two-photon polymerization (TPP) allows for producing arbitrarily complex geometries out of polymer at sub-micron resolution [23] while offering biocompatible resins [33]. Recently, TPP has gathered further attention as commercial printing systems began to deliver higher throughput [34] for additive manufacturing beyond the micrometer niche [35]. MN geometry and material, however, must be carefully selected, as it is critical that the MNs are strong enough to penetrate the skin's outermost barrier, the stratum corneum, resisting loads that may vary significantly with geometry [36], skin properties [37], insertion speed [38], and even local skin strain [39].

While TPP-made resin structures have had their mechanical properties extensively characterized [40], heterogeneity may arise in the micrometer scale [41], and the complex geometries and variability require not only finite element method (FEM) based simulations but also mechanical tests [42]. Most critically, TPP does not lend itself well to producing millimeter-sized monolithic systems [43]. Thus, the advancement of 3D-printed MNs is hindered by their device integration, as printing is commonly limited to solid base plates or devices without microfluidic interfacing due to technological challenges, such as micrometer-precise alignment [44].

Here, the precise direct 3D printing of hollow polymer MNs on silicon microfluidic chips by TPP for minimally invasive ISF extraction is presented. Through FEM simulations, the design was optimized for low hydraulic resistance while retaining mechanical robustness. A combination of mechanical failure (compression and shearing) and skin-surrogate insertion tests assessed the MN array's suitability for use in human skin. In vivo 72-hour biocompatibility and ISF extraction tests on two volunteers evaluated the platform's potential in continuous longitudinal health monitoring.

## 2. Results and discussion

*Microneedle patch design and fabrication*

The device's ultimate functions are penetration of the skin and extraction of ISF. To achieve this, the device interfaced 18 polymer hollow MNs with a Si-based microfluidic platform (Figure 1A) by additive manufacturing. The design was guided by a trade-off between maximizing the mechanical stability of the MN to penetrate the skin and minimizing flow resistance for ISF sampling. To reach the dermis, MNs were designed to be 1 mm long, and side holes were placed near their tip (Figure 1B, dimensions in Figure S1A, as discussed in Method Details). Additionally, the 18 MNs were arranged such that their connections to the ISF sensing chamber were of equal length (Figure S1B). This enables homogenous extraction from the targeted dermis area and similar residence time in the chip for the ISF extracted by any of the MNs, eliminating intrinsic temporal averaging, as desired to capture fast biomarker dynamics. To allow for long-term use, the MN array was designed as a wearable patch featuring a double-sided adhesive (featuring a hole that inscribes the MN array), enabling secure interfacing with the skin (Figure 1C). To allow the direct integration of sensors in the microfluidic reservoir, such as by sensor encapsulation [45], the patch employs an additional PDMS layer, which seals the microfluidic channels posterior to sensor integration and allows for the application of negative pressure for ISF extraction through the sensing element. This gives the patch great flexibility and integration potential, unlike monolithic 3D-printed designs [46].

Figure 1D shows the six innermost MNs in the array and a side view of one of the MN with its hollow core and twelve 30-µm side holes for ISF extraction (see Figure S1A for all the dimensions of the final MN design). FEM indicated a significant increase in the stress concentration factor when side holes were increased beyond 30 µm, as well as diminishing improvements in hydraulic performance (Figures S2A and S2B). A similar trend was observed for the side hole spacing, which was set at 40 µm to limit the reduction in the penetration depth of the side holes (Figure S2C). In contrast to conical MNs, the slender design of the TPP-made MNs allows for deeper penetrations while not significantly increasing the diameter of its body, reducing damage to the skin [29]. The inner diameter of the hollow MN was set at 50 µm, resulting from a compromise between resistance to ISF extraction and the additional stress compared to a non-hollow design (Figures S2D). Finally, the MN's base was fixed at 350 µm, as limited by the printer, to maximize adhesion between MN and chip. As this section of the MN does not penetrate the skin, the larger diameter does not cause further damage.

A clean-room-compatible process was developed to produce such a device (Figure 1E). By combining front and backside deep reactive ion etching, microfluidic chips were precisely made in a scalable manner before printing the MNs by TPP on top of the holes to create the array. As TPP allows for the fabrication of arbitrary geometries, such a process allows for rapid iteration over MN design and flexible material selection, as various polymers are available with distinct mechanical properties. The interfacing of the MNs with the substrate was also achieved with precise tilt and planar alignment, as shown by the low variance of MN height (i.e., relative error of -0.3%) and highly concentric (i.e., within 3 ± 2 µm) of the MN with respect to the microfluidic channels (Table 1), respectively. Precise control of the latter two is critical. Non-concentric needle-channel pairs can increase the hydraulic resistance of the device, diminishing ISF extraction. A high variance in the height may lead to MNs that are too short to reach the ISF-rich dermis and even the deterioration of mechanical stability due to MNs not properly attaching to the Si substrate (if prints start above the interface) or the reduction of the height of the flanged base (if far below the interface). Such tolerances are in line with or even outperform molded [47] and laser-ablated [48, 49] polymer-based MN devices reported in the literature, highlighting the precision achievable with the additive manufacturing process adopted here combined with established micromachining techniques.

*Mechanical testing of microneedles*

Due to their complex geometry and thick walls, FEM-based simulations were employed (Figures S2A-F) instead of commonly used analytical estimations [37] to locate the maximum stresses the MNs can be subjected to before failing under application-relevant compressive and shearing loads. Compression occurs during insertion into the skin, particularly when penetrating the stratum corneum (Figure 1B) [50]. If the MN cannot withstand such stresses, failure may lead to deformation during insertion or even fragmentation inside the skin, endangering subjects. The vicinity of the side holes was identified as a critical area for compressive failure (Figure 2A).

Stresses reached 14 MPa in the MN when a pressure of 3.16 MPa was simulated at the MN's tip (the stress required to pierce the stratum corneum [51]). Importantly, this maximum is still well below the material's 64.5 MPa ultimate stress. Maximum stresses at the side holes are in line with expectations due to their reduced cross-sectional area [52] and discontinuity-derived stress concentration [53]. Note that the MN's tip is also considered critical during the initial stages of penetration into the skin due to its minute cross-section.

Figure 2B shows the force versus displacement curve for an 18-MN array under such an application-relevant load when compressed against a stainless steel block (see Method Details). Forces increased until an average maximum of 7.39 ± 0.06 N (i.e., 411 ± 3 mN per MN) at 135 µm, approximately twice as much as usually required for skin penetration for similar MN geometries [54-56], highlighting the excellent mechanical properties achievable with 3D-printed polymer-based MNs. After that, the force decreases, and the MNs fracture at ca. 200 µm. A micrograph of the sample following the test (Figure 2C) indicates that fracture occurred near the side holes, as predicted by FEM (Figure 2A). Remarkably, the tested hollow MNs feature similar mechanical robustness to non-hollow thermoplastic MN arrays [47] despite stress concentrators derived from the hollow design [42].

During long-term use of the wearable patch, shearing stresses at the MN's base may come from everyday activities, such as dressing up or resting the arm on a surface. Such shearing of the MNs at their bases would render the patch useless, as the MN-microfluidic interface would no longer be airtight, and detached MNs would need to be individually removed from the skin. Thus, transversal forces near the MN's base were simulated by FEM (Figure 2D) and experimentally measured (Figure 2E). Forces rise sharply between 20 and 40 µm, plateauing at ca. 250 mN before dropping once the displacement reaches 80 μm. Such forces do not generate significant stresses in the MN (Figure 2D). Indeed, micrographs of the MN following the experiment revealed that the MNs detached from the substrate rather than fragmenting, and no permanent deformation was observed (Figure 2F).

*Skin surrogate penetration test*

Next, the penetration characteristics of the MNs were evaluated on a Parafilm® skin surrogate. Layered parafilm replicates the penetration depth and forces experimentally observed with human skin [57] despite its lack of heterogeneity (Figure 1B). Figure 3A shows the force versus displacement curves of an 18-MN array during 10 successive insertions and removals. During insertion, forces increase linearly with the penetration length due to the friction between the outer surface of the MNs and the skin surrogate. The force peaks at 3.1 ± 0.2 N (i.e., 172 ± 11 mN per MN) at the end of the insertion test, similar to the literature on real skin [37] and significantly lower than the experimentally determined failure force (safety factor of 2.4 ± 0.2, Figure 2B). While keeping the MNs inserted, the force drops to approximately 2 N as the surrogate relaxes, similar to skin [58]. Upon removal, forces decrease to zero and then become negative, indicating that the surrogate is no longer being compressed and that the friction forces resist the MNs' removal. The first (red line) and tenth (blue line) insertions did not differ significantly, and the array's re-insertion did not show any correlation with changes in maximum forces (R = -0.22, Figure S3). Importantly, no MN was lost during cycling, demonstrating the MN's robustness to insertion and that the attachment to the microfluidic platform is sufficiently strong.

*In vivo testing: Biocompatibility and ISF extraction*

Finally, tests were performed on two human volunteers to evaluate the material's biocompatibility. The volunteers had different ethnicities and BMIs (Table S1) – determining factors in successful MN insertion [59]. Following the devices' painless insertion (as confirmed by both subjects), the volunteers wore them for 72 hours, a desirable application time for continuous biomarker monitoring (e.g., cortisol [60, 61]). Subsequent photographic imaging of the insertion site following the wearable's removal showed quick recovery, with signs barely visible after one hour and not detectable after twelve hours (Figure 4A). Other reported MN perforations took up to two days to heal, likely due to their non-biocompatible materials and larger diameters [29]. Further, only a mild redness was perceivable, and the volunteers reported no pain, swelling, or itching, in contrast to other studies [62], indicating the excellent biocompatibility of the polymer-based hollow MNs. The micrographs of the arrays following the in-vivo test did not contain visible signs of deterioration to the MN's surface (Figure 4B), such as biofouling, a critical issue in MN-based biosensing [63].

As an early proof-of-concept, in vivo ISF extraction was tested by connecting the device to reservoirs that were constantly evacuated by a microfluidic pump down to 0.3 bara (Figure 4C). After one hour, approximately 1 µl of ISF was collected in the tube. As a result, we demonstrated the successful extraction of ISF with 3D-printed polymer hollow MNs on a microfluidic chip.

## 3.  Conclusion

The precise interfacing of 3D-printed MNs with Si-based microfluidic chips was developed, resulting in a wearable patch for ISF extraction. The hollow polymer MNs had minimal height variability and were accurately positioned on the microfluidic chip, demonstrating the compatibility of additive manufacturing and Si-based cleanroom processes. The so-produced MN arrays feature excellent mechanical resistance to compression and shearing, being capable of withstanding at least ten consecutive penetrations into skin surrogates. When the device was tested for real-world application in volunteers for several days (72 h), no pain during insertion, absence of skin irritation, and rapid wound recovery revealed the outstanding applicability and biocompatibility of the polymer-based hollow MNs. The extraction of ISF from human subjects provides a proof of concept for wearable MN-based longitudinal health monitoring.

This versatile platform is a promising solution to decouple sampling and sensing in MN systems for the analytical characterization of biomarkers in ISF. The adopted TPP production enables flexible changes in MN geometry, allowing the design to be adapted to skin conditions in the region to be explored. The well-established Si microfluidics also lends itself to iterative changes and may be adapted to passive extraction modes (e.g., capillary pumps). Such a platform can incorporate many sensing materials and extend continuous health monitoring to many analytes, replicating the disruption brought by glucose monitoring.

## 4. Method Details

*FEA-informed design of hollow microneedles*

The dimensions of the final MN design are depicted in Figure S1. MNs were simulated by FEM (COMSOL Multiphysics, COMSOL AB, Sweden). The MN consists of three main sections: a sharp conical tip, a main body section with a slight wall angle, and a flanged base (Figure S1A). Side-hole spacing was varied between 35 and 80 µm, while their diameter was between 10 and 35 µm (both 5 µm steps). The height of the MN was fixed at 1 000 μm, as dictated by the application. The inner diameter was varied between 30 and 90 μm in 10 μm steps. Mesh elements were kept between 1.2 and 3.6 μm to keep the discretization errors in the vicinity of the side holes below 0.2% (Figure S2G). Finally, The geometry was discretized using quadratic serendipity elements [65], and the model was solved using the indirect GMRES solver [66]. Given the MN's symmetry, only one-quarter of the geometry is simulated (Figure S2H). The dependence of the stresses experienced by the MNs on geometric features was derived for compressive and bending loads. The model was repeatedly solved for each parameter value through parametric sweeping of one variable at a time. The IP-S resin was modeled as a linear elastic material with Young's module of 1.6 GPa and a compressive strength of 64.5 MPa [40, 67]. For compression simulations, symmetry boundary conditions are imposed on the model's faceted sides; for shearing, a symmetric boundary condition is imposed on the side normal to the force and an asymmetric one on the orthogonal side. A rigid boundary condition is put on the base of the MN, where it's interfaced with the Si substrate. The load is uniformly distributed over the MN's tip's surface area.

*Design and fabrication of microfluidic chips*

The dimensions of the microfluidic channels are shown in Figure S1B. Microfluidic chips were fabricated from single-side polished, 500-μm-thick, 100 mm boron-doped silicon wafers with a (100) orientation (Prolog Semicor LTD, Ukraine). They consisted of a sensing chamber attached to a microfluidic network, which connected it to equidistant through holes on top of which MNs were printed (Figure S1B). A schematic representation of the process can be seen in Figure 1E. To produce the holes that connect the MNs to the microfluidic chip, a 6-μm-thick layer of AZ 4562 photoresist (Clariant AG, Switzerland) was spin-coated (4 000 rpm, Süss MicroTec, Germany) on the front (non-polished) side of the wafer and used as an etching mask for a 400-μm-deep modified 3-step Bosch process [68]. The hardened photoresist was then removed by plasma ashing (PlasmaPro 100 Estrelas, Oxford Instruments plc, UK), and the wafer was cleaned in acetone and IPA (both >98%, Sigma Aldrich, USA) ultrasonic baths. Then, to form the microfluidic channels, these processes were repeated on the wafer's back (polished) side with a 5-μm-thick photoresist layer (6 000 rpm) to produce a 150-μm-deep etch. All photolithography steps were performed on an MA/BA6 Maskaligner (Süss MicroTec, Germany) with a 400 mJ cm$^{-2}$ dose of 405 nm light. Masks were produced by a DWL 2000 (Heidelberg Instruments Nano AG, Germany). Patterns were developed by immersing the wafer in a 17% solution of Microposit Developer 351 (micro resist technology GmbH, Germany) in water. Wafers were cut with a dicing saw (DAD 3221, DISCO Corporation, Japan) into 25-mm square dies. 250-μm notches were made to allow the individual devices on each die to be separated following the array fabrication.

*3D printing of hollow microneedles*

MNs with the design specified in Figure S1A were printed on the front side of the wafer by two-photon polymerization by dip-in lithography [69] using a Nanoscribe GT2 printer (Nanoscribe GmbH, Germany) equipped with a 25x-magnification objective (Carl Zeiss AG, Germany). The objective was immersed in IP-S (Nanoscribe GmbH, Germany) and focused a 50-mW laser of 780 nm wavelength. The layer height was set to 1.5 μm, and the hatching distance between voxels to 0.7 μm. Most of the structure employed a 100 000 μm s$^{-1}$ scan speed, but the first 15 layers employed 20 000 μ s$^{-1}$ to increase the degree of polymerization [70] at the MN-chip interface. Three separate printing blocks were used to overcome the working distance limitations; an overlap of 10 μm between blocks was used to ensure interface stability.

MNs were developed in mr-Dev 600 (micro resist technology GmbH, Germany) for 10 min before being washed by immersion in IPA for 15 min. Following development, the devices were heat-treated at 190 °C on a hotplate (Harry Gestigkeit GmbH, Germany) to ensure no unpolymerized resist remained at the MN-chip interface and further improved adhesion. In-plane alignment was performed with a custom-made GWL (Nanoscribe GmbH, Germany) code, and alignment marks were imprinted during etching. Out-of-plane misalignment (i.e., the non-orthogonality between the laser and the chip) was compensated by a Python script that fits the best slope based on a 24-point interface measurement.

Following printing, dies were broken along the diced notches. Devices meant for microfluidic testing were further processed to connect them to a microfluidic pump. First, the microfluidic device and a 500-µm-thick PDMS sheet (Limitless Shielding, UK) were placed inside an oxygen plasma chamber (100 W, Diener Electronic GmbH, Germany) and then brought into contact to form a covalent bond at the PDMS-device interface. A punch-hole 0.5 mm in diameter was then made to allow access to the device's microfluidic network. Finally, a fitting made from a 3-mm-thick PDMS sheet was bound on top of the punch-hole by a similar oxygen plasma process.

*MN imaging*

Secondary electron (SE) SEM micrographs were obtained in a field emission TFS Magellan 400 (Thermo Fisher Scientific, USA) equipped with an Everhart-Thornley detector. SE imaging was carried out at a 3 kV acceleration voltage. Samples were sputtered with a thin (i.e., ca. 5 nm) layer of Pd/Pt before imaging. Optical micrographs were obtained with a VHX-7000 digital microscope (Keyence Corporation, Japan) in full-ring reflection illumination. Scale bars were not corrected for tilt and thus represent the projected lengths.

*MN mechanical testing*

Compression tests were performed on MN arrays employing a 3D-printed holder attached to an Instron 5848 Microtester (Instron, USA) that compressed the arrays against a stainless steel block at 0.5 mm s$^{-1}$ until a 250 µm displacement was reached.

Shear tests were performed on single MNs with a 5-axis CNC machine (5XM400, 5AXISMAKER, UK) adapted with a force torque sensor (MiniOne Pro, Bota Systems AG, Switzerland) setup. Data was acquired in 5 µm increments. Steps were taken every three seconds, allowing the sensor to measure thrice for each step.

Insertions tests were carried out similarly to compression tests but substituting the stainless steel block with a holder containing eight layers (127 µm per layer) of Parafilm M® (PF, Bemis, USA) supported by expanded polyethylene as a skin surrogate [57]. The arrays were inserted 1 mm below the PF's surface at 0.5 mm s$^{-1}$, held in place for 10 s [58], and then retracted. Arrays were always inserted into previously unperforated PF skin surrogates.

*In-vivo ISF extraction and biocompatibility test*

Wearable patches were inserted into the forearms of two healthy male subjects (Table S1) at 3.4 m s$^{-1}$ [38] with a purpose-built device. A microfluidic tube (1/16-inch outer diameter) is then inserted into the PDMS fitting and connected to a microfluidic pump (LineUp Push-Pull, Fluigent, France). The pressure was then reduced to -700 mbarg in -100 mbarg steps and maintained for one hour. Following this period, a previously weighted collection tube is removed, and its weight variation is determined by a precision scale (AB1135-S/FACT, Mettler Toledo, USA). The volume of collected ISF is then estimated by assuming a specific mass of 1 000 kg m$^{-3}$ [71]. Biocompatibility tests were performed by keeping the so-inserted MN arrays in the male subject for 72 hours. In-vivo tests received a waiver of approval from the responsible ETH Zurich Ethics Commission (IRB No. IRB00007709).

## CRediT authorship contribution statement

**Tiago Elias Abi-Ramia Silva:** Writing – original draft, Writing – review & editing, Visualization, Investigation, Data curation, Methodology, Validation, Conceptualization. **Stephan Kohler:** Writing – review & editing, Visualization, Investigation, Methodology, Conceptualization. **Nicolas Bartzsch:** Writing – review & editing, Visualization, Investigation, Methodology, Conceptualization. **Felix Beuschlein:** Writing – review & editing,  Methodology, Conceptualization. **Andreas T. Güntner:** Writing – review & editing, Visualization, Supervision, Investigation, Funding acquisition, Conceptualization.

## Declaration of Competing Interest

The authors declare that they have no known competing financial interests or personal relationships that could have appeared to influence the work reported in this paper.

## Data and code availability

The data supporting this study's findings are available from the corresponding author upon reasonable request.

## Acknowledgment

This study was primarily funded by the Swiss State Secretariat for Education, Research, and Innovation (SERI) under contract number MB22.00041 (ERC-STG-21

"HEALTHSENSE"). FB acknowledges funding by the Clinical Research Priority Program of the University of Zurich for the CRPP HYRENE.

The authors thank Prof. Dr. C. Hierold and F. Püntner for their support with mechanical characterization, V. Gantenbein for his support with 3D printing, D. Scheiwiller for his support with process development, and Dr. H. Shin and T. Raicevic for their support with imaging (all ETH Zürich).

## Quantification and statistical analysis

Statistical analyses were performed using measurements from ImageJ and statistical functions from Microsoft Excel. Data are expressed as mean ± standard deviation. Variables are considered weakly correlated when Pearson's correlation coefficient equals or is below 0.29. Statistical details of experiments can be found in the figure legends.

## 5. Figures and Tables

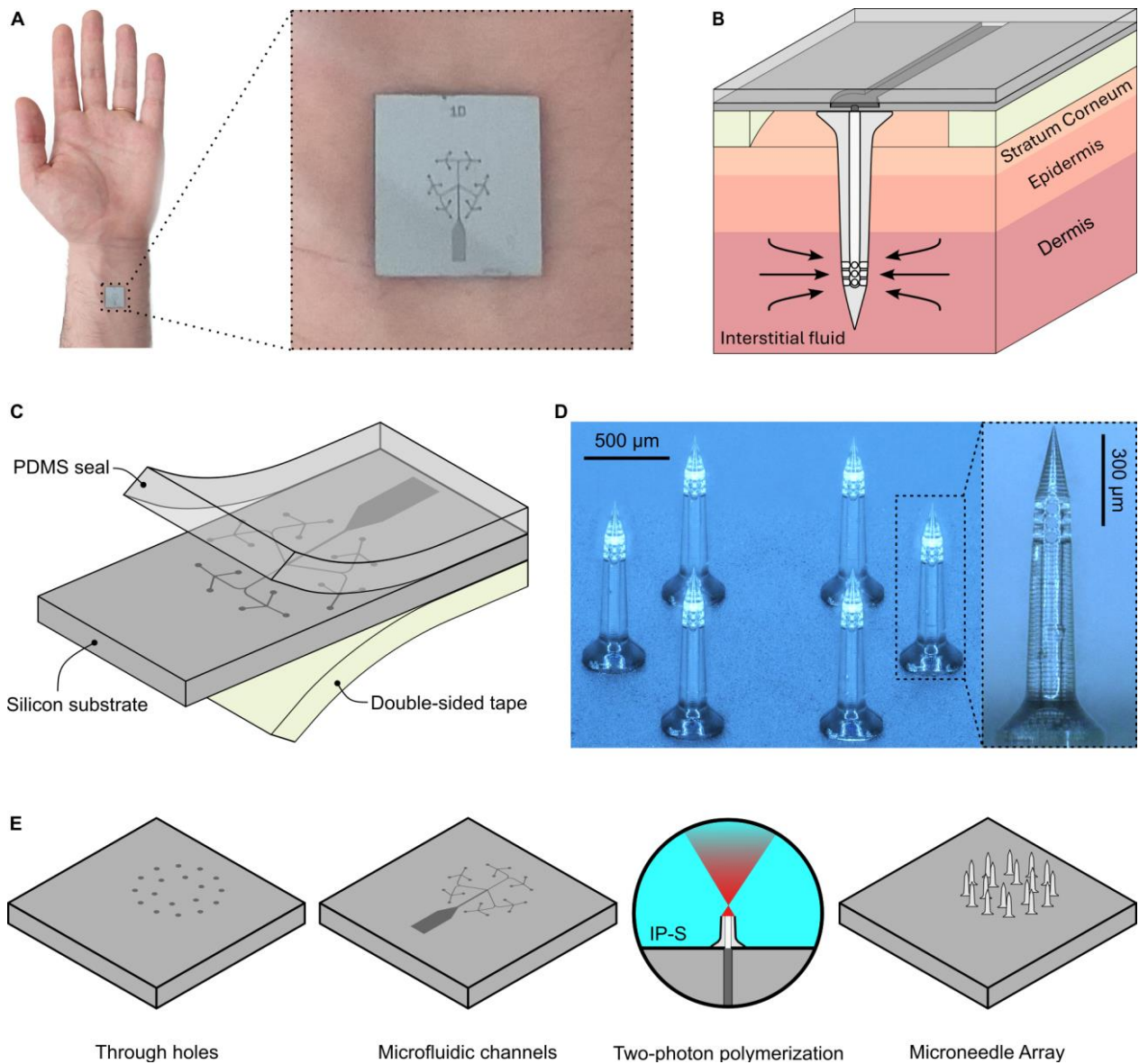

**Figure 1. Wearable MN patch design and fabrication.**

(A) A MN array inserted into a subject's forearm. Note that the PDMS seal was removed for better visualization.

(B) ISF sampling concept: the hollow MNs' side holes reach the skin's dermis, where they access the biomarker-rich ISF. The ISF is transported to the sensing chamber through the hollow MNs and the microfluidic platform's channels.

(C) View of the designed patch with the layers that are added following printing.

(D) Optical micrograph of an array's six-MN cluster and a detailed view of a single MN.

(E) Si-based microfluidic chips are fabricated through front- and back-side DRIE processes that produce the through holes and the microfluidic channels. The resulting dies are then used as a printing substrate upon which IP-S is deposited and polymerized by TPP, producing 18 MNs per patch.

See also Figure S1.

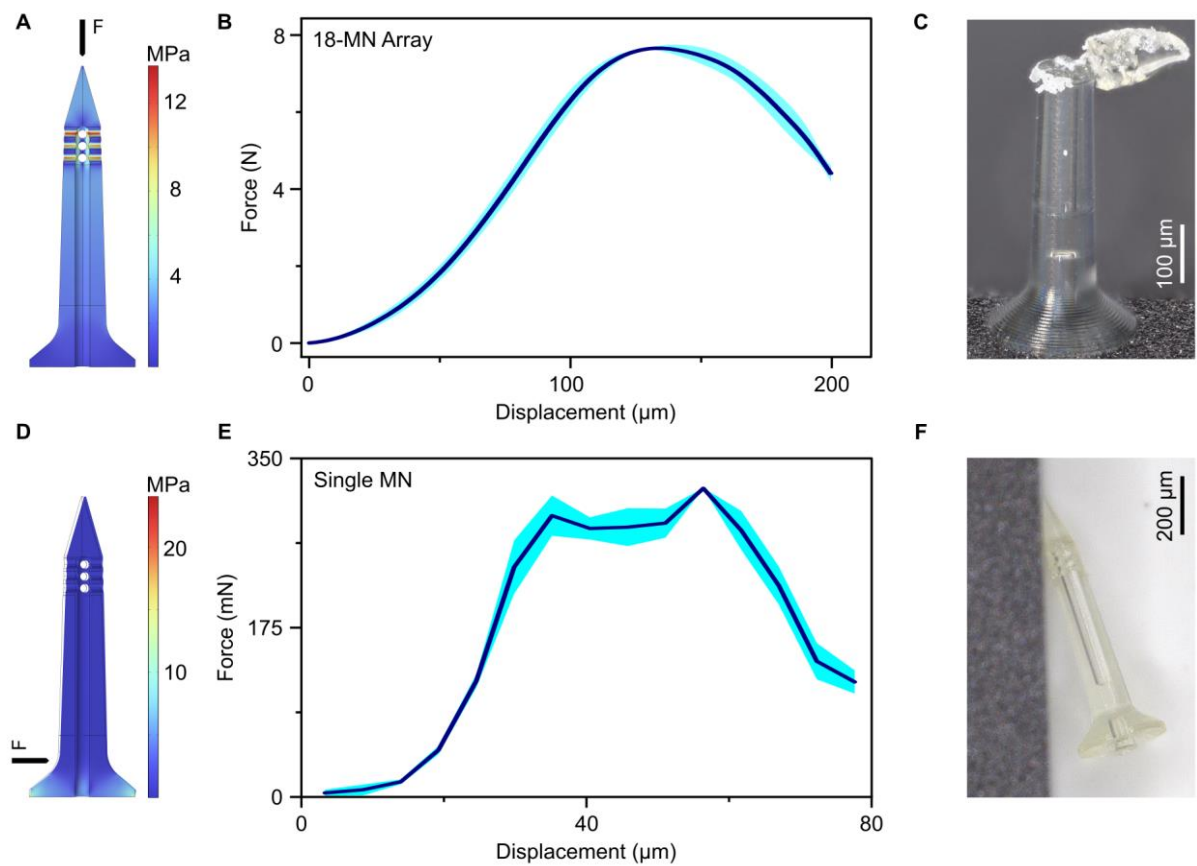

**Figure 2. MN mechanical characterization and failure analysis.**

(A) FEM simulation of the MN under a compressive load.

(B) Constant speed compression test performed on 18-MN arrays. The curve is shown as an average (dark blue) and the error band indicates one standard deviation (light blue) (n = 3).

(C) Micrograph of an MN following failure under compressive load.

(D) FEM simulation of the MN under a shear load.

(E) Step-wise shear test performed on a single MN. The curve is shown as an average (dark blue) and the error band indicates one standard deviation (light blue) (n = 3).

(F) Micrograph of an MN following detachment from the microfluidic platform under shearing load.

See also Figure S2.

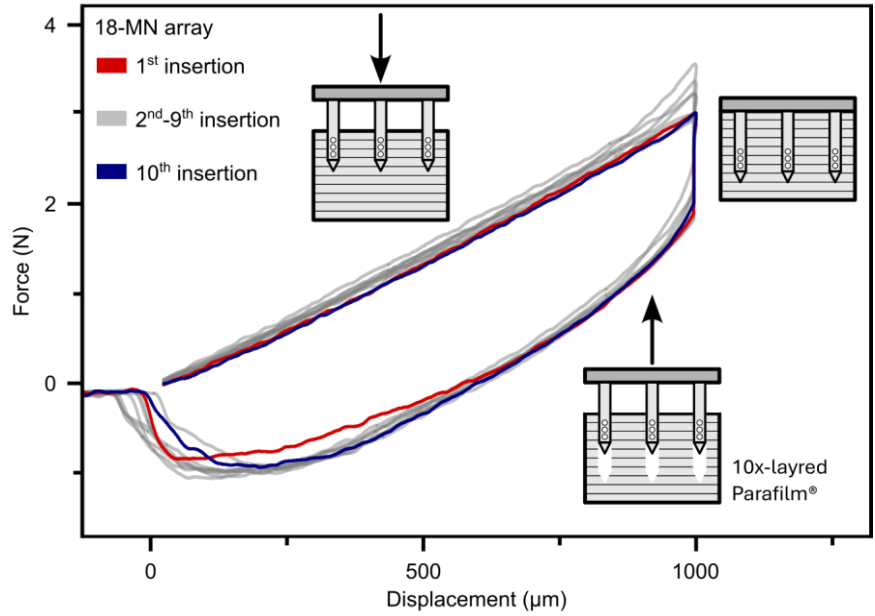

**Figure 3. Skin surrogate penetration.**

MN insertion force versus displacement over ten successive insertions into unperforated areas of a Parafilm® skin surrogate.

See also Figure S3.

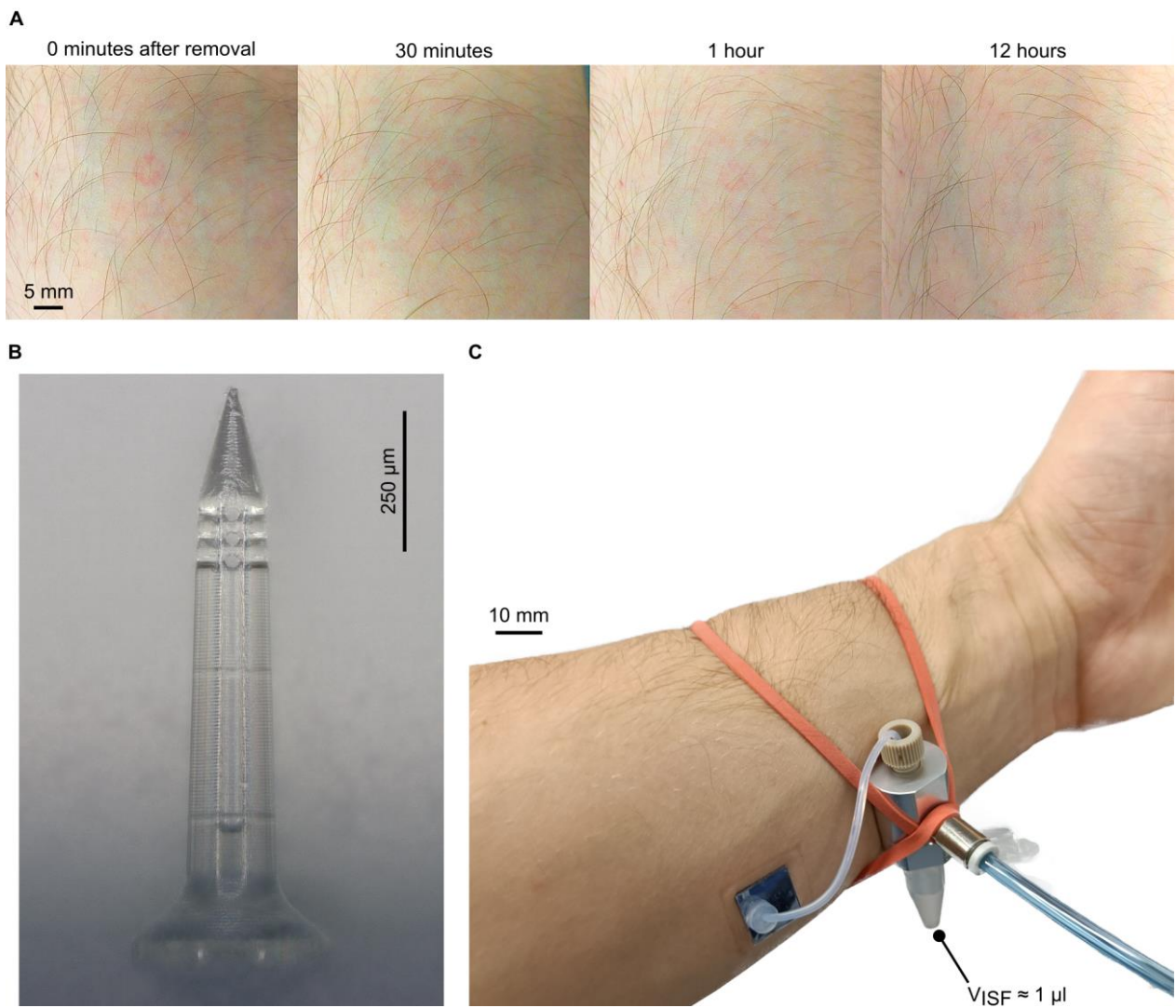

**Figure 4. In vivo and ISF extraction tests.**

(A) Pictures of the subject's skin immediately following the three-day biocompatibility test and its recovery after 30 minutes, 1 hour, and 12 hours.

(B) Optical micrograph of a used MN after 72-h in vivo test.

(C) ISF extraction being performed by the wearable patch connected to a microfluidic pump (not shown).

See also Table S1.

**Table 1.** Printing precision assessment of 18-needle MN arrays performed with SEM (height) and optical microscopy (concentricity). See also Figure S1.

| Parameter (µ ± σ) | MN | Nominal | Relative Error |
|---|---|---|---|
| Height [a] | 997 ± 2 µm | 1 000 µm | -0.3% |
| Concentricity | 3 ± 2 µm | 0 µm | - |

[a] 15 µm is added to the measured height as the printer is set to start below the surface by this amount.

Tuning of Cellulose Nanocrystals Reinforced Polymer Nanocomposites, Small 19(3) (2023) 2202470.

[68] M.S. Gerlt, N.F. Läubli, M. Manser, B.J. Nelson, J. Dual, Reduced Etch Lag and High Aspect Ratios by Deep Reactive Ion Etching (DRIE), Micromachines 12(5) (2021) 542.

[69] S. Maruo, O. Nakamura, S. Kawata, Three-dimensional microfabrication with two-photon-absorbed photopolymerization, Opt. Lett. 22(2) (1997) 132-134.

[70] Y. Yuan, L. Chen, Z. Shi, J. Chen, Micro/Nanoarchitectonics of 3D Printed Scaffolds with Excellent Biocompatibility Prepared Using Femtosecond Laser Two-Photon Polymerization for Tissue Engineering Applications, Nanomaterials 12(3) (2022) 391.

[71] W. Yao, Z. Shen, G. Ding, Simulation of Interstitial Fluid Flow in Ligaments: Comparison among Stokes, Brinkman and Darcy Models, International Journal of Biological Sciences 9(10) (2013) 1050-1056.

# Supplementary Information

# 3D printing by two-photon polymerization of hollow microneedles for interstitial fluid extraction

Tiago Elias Abi-Ramia Silva,[1] Stephan Kohler,[1] Nicolas Bartzsch,[1]

Felix Beuschlein,[2,3,4] and Andreas T. Güntner[1,2,*]

[1]*Human-centered Sensing Laboratory, Department of Mechanical and Processing Engineering, ETH Zurich, CH-8092 Zurich, Switzerland*

[2]*Department of Endocrinology, Diabetology, and Clinical Nutrition, University Hospital Zurich (USZ) and University of Zürich (UZH), CH-8091 Zurich, Switzerland*

[3]*Medizinische Klinik und Poliklinik IV, Klinikum der Universität, Ludwig-Maximilians-Universität, Munich, Germany*

[4]*The LOOP Zurich – Medical Research Center, Zurich, Switzerland*

*Corresponding author.*
E-mail address: andregue@ethz.ch

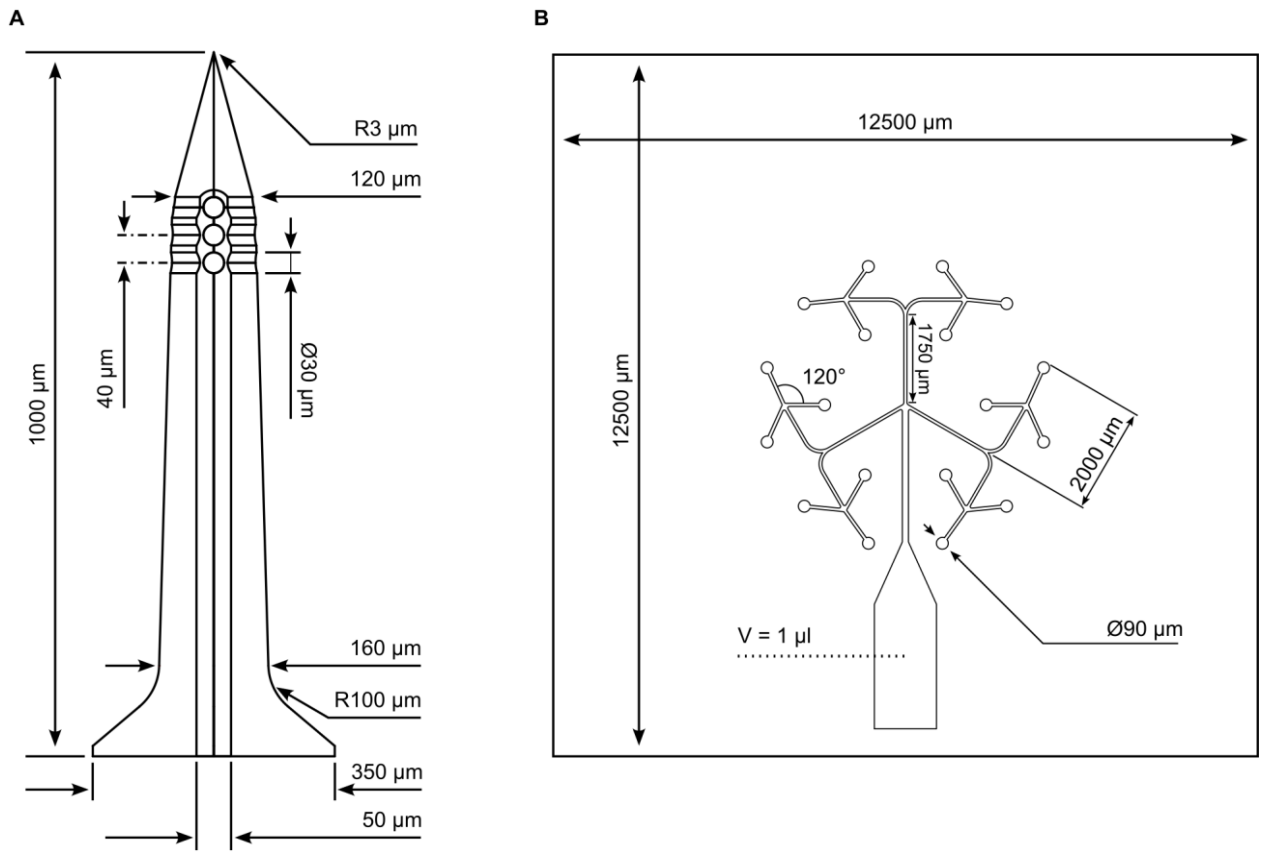

**Figure S1. Final microneedle design, related to Figure 1.**

(A) Key dimensions of the produced MN designed. Values as guided by the FEM analysis in Figure S2.

(B) Key dimensions of the produced microfluidic chip.

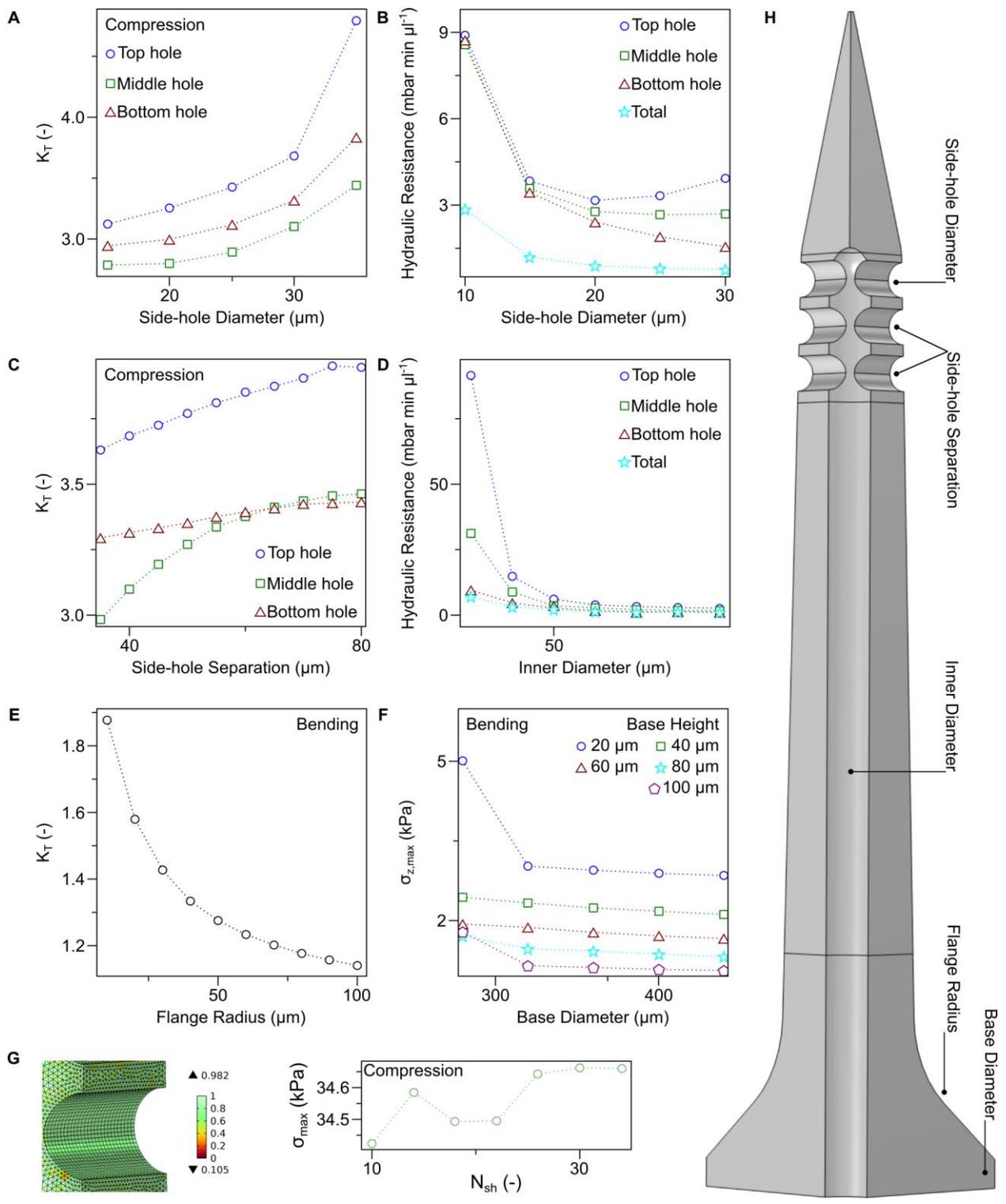

**Figure S2. FEM-guided microneedle design, related to Figure 2.**

(A) Stress concentration factor ($K_T$) in the vicinity of the side holes as a function of their diameter for a compressive load.

(B) Resistance to ISF flow through the MN as a function of the side hole diameter.

(C) Stress concentration factors in the vicinity of the side holes as a function of their spacing for a compressive load.

(D) Resistance to ISF flow through the MN as a function of its inner diameter.

(E) Stress concentration factors in the vicinity of the base as a function of the flange's fillet radius for a bending load.

(F) Maximum out-of-surface stress ($\sigma_{z,max}$) at the MN-chip interface as a function of the base's diameter during bending for different base heights.

(G) Side hole mesh element skewness and maximum stress convergence under compression study.

(H) FEM geometry used for the mechanical studies (compression, bending, and shearing).

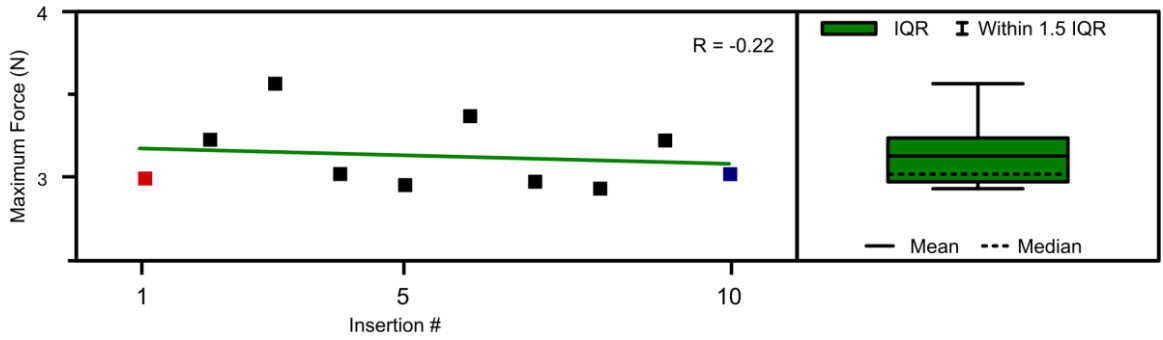

**Figure S3. Statistics on insertion forces, related to Figure 3.**

Maximum force versus insertion number and the box plot of the data.

Table S1. Self-reported physiological data of the volunteers, related to Figure 4.

| Subject | Age [years] | Sex | Weight [kg] | Height [m] | Race |
|---|---|---|---|---|---|
| #1 | 23 | Male | 78 | 1.79 | White |
| #2 | 27 | Male | 64 | 1.70 | Multiracial |